\documentclass[times,twocolumn,authoryear]{revtex4-2}

\usepackage{framed,multirow}
\usepackage[pdftex]{graphicx}
\graphicspath{{figures/}}
\usepackage[T2A]{fontenc}
\usepackage[utf8]{inputenc}

\usepackage[english]{babel}
\usepackage{amssymb}
\usepackage{latexsym}

\usepackage{amsmath}
\usepackage{cmap}
\usepackage{multirow}
\usepackage{layouts}

\usepackage{caption}
\usepackage{placeins}

\usepackage[usenames]{color}

\usepackage{url}
\usepackage[dvipsnames]{xcolor}
\definecolor{newcolor}{rgb}{.8,.349,.1}

\usepackage[colorlinks=true,urlcolor=blue,citecolor=red,linkcolor=red,bookmarks=true]{hyperref}

\newcommand{\Mark}[1]{{#1}}

\begin{document}

		\title{Numerical simulations  of a continuously injected relativistic electron beam relaxation into a plasma with large-scale density gradients}%
		
		 \author{V. V. Annenkov (В.В. Анненков)}
		\affiliation{Budker Institute of Nuclear Physics SB RAS, 630090, Novosibirsk, Russia}
		\affiliation{Novosibirsk State University, 630090, Novosibirsk, Russia}

		\author{E. P. Volchok (Е.П. Волчок)}
		\affiliation{Budker Institute of Nuclear Physics SB RAS, 630090, Novosibirsk, Russia}
		
		\affiliation{Novosibirsk State University, 630090, Novosibirsk, Russia}
%
%

		\begin{abstract}
			
			In this paper the influence of large-scale decreasing and increasing gradients of the density of magnetized plasma on the relaxation process of a continuously injected relativistic electron beam with an energy of 660 keV ($v_b=0.9c$) and a pitch-angle distribution is studied using particle-in-cell numerical simulations.  It is found that for the selected parameters in the case of a smoothly decreasing gradient and in a homogeneous plasma the formation of spatially limited plasma oscillations of large amplitude occurs. In such cases, modulation instability develops and a long-wave longitudinal modulation of the ion density is formed.  In addition, the large amplitude of plasma waves accelerates plasma electrons to energies on the order of the beam energy. In the case of increasing and sharply decreasing gradients, a significant decrease in the amplitude of plasma oscillations and the formation of a turbulent ion density spectrum are observed. The possibility of acceleration of beam electrons to energies more than 2 times higher than the initial energy of the beam particles is also demonstrated.  This process takes place not only during beam propagation in growing plasma density, but also in homogeneous plasma due to interaction of beam particles with plasma oscillations of large amplitude. 	
			\end{abstract}

	\maketitle	
	\section{Introduction}\label{sec:Int}

	Interaction of electron beams with plasma takes place in a large number of physical systems. One of the most important processes during the electron beam relaxation is a two-stream instability, as a result of which beam particles can drive plasma oscillations at the local plasma frequency and with a phase velocity equal to the velocity of the beam. In homogeneous plasma, such oscillations are potential and can not escape plasma in the form of electromagnetic (EM) radiation. However, the existence of density gradients, the mechanisms of linear mode conversion  \citep{Sakai2005,Malaspina2012,Timofeev2015}, the conversion of electrostatic Langmuir turbulence \citep{Ziebell2015},  the backward waves formation \citep{Ginzburg1958,Zheleznyakov1968} and the presence  of counter propagating waves in plasma \citep{Ganse2012,Annenkov2018,Annenkov2020} transform plasma oscillations into EM radiation at harmonics of plasma frequency.  In heliophysics, such processes appear as different bursts in radio and THz ranges. The cited papers do not cover the entire range of works on the topic of research of radiation generation mechanisms. More details can be found in the references in the cited works, as well as in the reviews \cite{Wild1963,Bastian1998,Reid2014}. In a magnetized plasma, in addition to oscillations at the plasma frequency, electron beams can excite a wide range of different oscillations with complex dispersion properties.

	Under laboratory conditions, plasma mechanisms for EM radiation generation are currently studied as promising sources of high-power narrowband THz radiation \citep{Arzhannikov2020}. Also injection of non-relativistic electron beams into a gas is a perspective technique for creating a target plasma in open magnetic traps for plasma confinement, while beam-excited transversely polarized whistlers can lead to the formation of electron populations accelerated to energies an order of magnitude greater than the initial energy of the beam particles \citep{Soldatkina2021,Timofeev2021}.

	Density gradients are common for all beam-plasma systems. These can be a sufficiently small-scale inhomogeneity of the dimension less than or on the order of plasma wavelength, induced by propagating sound waves or developing modulation instability. During propagation through the solar atmosphere, beams of fast electrons interact with  a large-scale gradients of plasma density \citep{Aschwanden2006}. In plasma of laboratory facilities, such gradients can arise due to peculiarities of gas pumping and target plasma creation. Additionally, different random density fluctuations can take place \citep{Bruno2013}, due to  all previous history of a plasma system. The presence of density gradients can essentially affect  the two-stream instability development and subsequent processes of plasma heating, particle acceleration and EM emission generation.

	Quasilinear relaxation of an electron beam with a small plasma inhomogeneity in the framework of the weak turbulence theory was considered in the works \textcite{Ryutov1969,Breizman1972}, this approach was extended to the case of large amplitude fluctuations in \textcite{Voshchepynets2013}. Different  density gradients are considered in  \textcite{Kontar2001,Kontar2001a,Kontar2002}. Kraft and Volokitin \citep{Krafft2013,Krafft2015,Krafft2019,Volokitin2018,Volokitin2020,Krafft2020} actively study influence of random plasma density inhomogeneities on plasma oscillations and solar radio bursts generation in Hamiltonian approach whereby the dynamics of the background plasma particles, Langmuir and ion-acoustic wave fields are described by the Zakharov equations. Main prediction of quasilinear theory and hamilton model are confirmed by particle-in-cell (PIC) simulations \citep{Thurgood2016}. In \textcite{Krafft2021}, comparison of density inhomogeneities impact on plasma oscillations in models based on Zakharov equation and geometric optics are presented. Similarity of these theories predictions has been demonstrated.

	In works \textcite{Shalaby2018,Shalaby2020} the impact of plasma density inhomogeneities on the growth rate of beam-plasma instability was investigated using one-dimensional PIC simulations. Relaxation of a power-law non-thermal electron population in a collisional inhomogeneous plasma is investigated in the context of electron acceleration research and interpretation of hard X-ray spectra in solar flares in \textcite{Kontar2012}. Also the study of electron acceleration during propagation of an electron beam in plasma with a positive density gradient is studied using 1D modeling in paper \textcite{Kudryavtsev2019}. In work \textcite{Krafft2021a}, 2D PIC simulations of an electron beam relaxation in plasma with random density fluctuations are carried out. It is shown that in this case the reverse plasma oscillations, which are necessary for the generation of EM radiation through the three-wave process $L+L'\rightarrow T_{2\omega_p}$, occur earlier in the system. This leads to higher emission level at the second harmonic of the plasma frequency. The theoretical model for generation of harmonic emissions of Type III solar radio bursts in strongly inhomogeneous plasmas  is proposed in \cite{Tkachenko2021}. In work \cite{Vatagin2021}, the evolution of the electron distribution function and the spectral energy density of Langmuir turbulence with a pulsed injection of electrons in an inhomogeneous flare plasma is studied.

	\citet{Pechhacker2012} investigated the effect of electron beam pitch angle and parabolic density gradient on solar type III radio bursts using 1.5D PIC simulations. Effects of similar plasma density gradients  but in the case of a large magnetic field ($\Omega_e>\omega_p$) are considered in 2D PIC simulations by \citet{Yao2020}. In paper \cite{Pechhacker2014}, the appearance of accelerated electrons in plasma with exponential density gradients was investigated using 3D PIC simulation. The effect of an exponentially growing plasma profile on the growth rate of beam instability is also considered \citep{Bret2006} in the frameworks of Wentzel-Kramers-Brillouin approximation for inertial thermonuclear fusion  problem.

	Let us formulate basic effects of density inhomogeneities affecting the process of a beam relaxation:
	\begin{itemize}
		\item While a Langmuir wave propagating towards the growing density gradient, the longitudinal wave number $k_\parallel$ decreases. As a result, the phase speed $v_{ph}=\omega/k_\parallel$ increases. This allows waves to resonantly exchange energy with particles of a higher speed, creating high-velocity particle jets. This process is the main mechanism of appearing particles with energies higher than the beam energy in the quasilinear relaxation theory.
		
		\item If the plasma density changes significantly  on the wavelength scale of the beam-excited plasma oscillations (resulting in a change in the local plasma frequency), it becomes difficult to excite plasma oscillations in this region, despite a possible local increase in the relative density of the beam. As a result, there is a local disruption of instability.
		\item Random density fluctuations can stimulate the formation of backward plasma waves and EM radiation at harmonics.
		\item Specially prepared density gradients of plasma make it possible to directly convert plasma waves into EM radiation.
	\end{itemize}

	In cited above works, both analytical as well as based on numerical simulations, a  beam-plasma system is considered infinite: random fluctuations of density have periodic structure, periodic boundary conditions in the beam propagation direction are superimposed on a computational domain in simulations. An analogue problem formulation is used even in the systems with large-scale density gradients, which required specific modifications of a computational domain. As an example, in  \citet{Pechhacker2014}, particle equations are solved with the periodic boundary conditions but field equations are with open ones. In works \citet{Pechhacker2012,Yao2020} a descending density profile transforms to a rising density profile, in order to provide periodicity and match of system edges. For more realistic formulation, instead of an infinite beam, a beam limited in the longitudinal direction, with a length several times smaller than the computational domain, is used.

	The duration of many plasma processes of interest  involving electron beams many times exceeds the time scales of hundreds and thousands $\omega_p^{-1}$ which are typically considered in PIC simulations. One examples of such systems is  type III solar radio bursts \citep{Reid2018}. In modern laboratory experiments on a beam-plasma discharge, the duration of beam injection and generation of radiation at harmonics of plasma frequency has a microsecond scale, with a characteristic period of plasma oscillations of the order of picoseconds \citep{Arzhannikov2020}. In this case, for the processes at the specific finite plasma region (e.g. a source of radiation) to be considered, the model with open boundary conditions and continuously injected beams is physically more correct. The basic difference between such a model and the infinite plasma model is the constant inflow of <<fresh>> beam particles with initial energy into the system. This leads to qualitative differences in the results of beam relaxation in plasma. For example, due to the presence of a constant source of energy, the plasma oscillations excited by the beam reach a much higher amplitude. The interaction of such oscillations with the incoming beam particles leads to the appearance of particles with an energy many times greater than the initial energy of the beam particles. At the same time, this does not require the presence of any density gradients (as in the quasilinear theory) or allowance for ion dynamics. Also, for sufficiently dense beams, another scenario of ion evolution is observed. Instead of excitation of ion-acoustic oscillations, the formation of longitudinal density modulation under the action of the ponderomotive force of the high-frequency field can take place.  Many features of the model with continuously injected beams are discussed in papers \cite{Annenkov2019,Annenkov2020} and references therein.

	Let us consider the currently available results of studying the effect of large-scale density gradients on the beam relaxation process in a model with open boundary conditions. Two unusual effects have been observed in experiments on the injection of a thin ($r<c/\omega_p$) sub-relativistic (energy of the order of 20 keV) electron beam into a neutral gas at the GDT open magnetic trap (BINP SB RAS) \citep{Soldatkina2021}. The first effect was the formation of plasma in a volume much larger than the area occupied by the beam, and the second one was the appearance in the system of electrons accelerated to energies of the order of 300 keV. To interpret the observed effects, 2D3V PIC simulations of the electron beam injection into the region near the input magnetic mirror have been carried out, taking into account the growing longitudinal and decreasing transverse density gradients and magnetic field curvature \citep{Timofeev2021}.  Numerical experiments have demonstrated the acceleration of electrons in the region of beam-plasma turbulence up to a value of the order of 100 keV in one act of interaction with a plasma wave. However, such particles became transient and had to immediately leave the magnetic trap. But the simulation data also showed the possibility of excitation of transversely polarized whistlers in the region inside the magnetic trap. Such waves are capable of many times accelerating trapped electrons without bringing them into the loss cone. To confirm the hypothesis about such a mechanism of particle acceleration in the trap, additional real experiments were carried out, which showed its plausibility.

	In work \cite{Annenkov2019}, generation of radiation during injection into a plasma of a 100 keV electron beam with a transverse dimension of the order of the plasma wavelength has been studied. The results obtained made it possible to explain the experimentally observed level of EM emission at plasonics \citep{Burdakov2013}. However, in the simulation of beam injection into a homogeneous plasma, the region of relaxation and generation of radiation turned out to be localized at a distance of about 5 cm from the region of beam injection, while in the experiment intense radiation was observed at a distance of about 80 cm from the region of beam injection into the plasma. To explain this fact, simulations were carried out with considering the large-scale positive density gradient (an increase in density by a factor of 2 over a length of 1 meter) present in the experiment. However, taking into account such a gradient did not have a sufficient stabilizing effect, and the relaxation region shifted only up to 8 cm. Also, the modulation of the ion density formed during beam relaxation did not lead to a sufficient disruption of the instability. However, when both the gradient and the turbulent plasma density are taken into account, no development of beam-plasma instability was observed over a length of about 15 cm.

	During acceleration at the CME-driven shock or in magnetic reconnection events, electron beams with relativistic velocities can be generated \citep{Klein2022}. Propagating in inhomogeneous solar coronal plasma, such beams lead to the generation of radiation in the radio band at plasma frequency harmonics. Depending on the direction of propagation relative to the surface of the Sun, these beams interact with the plasma with increasing (upward propagation) or decreasing (downward propagation) density.  In this paper, without pretending to consider the question comprehensively, we restrict ourselves to the relativistic beam (the speed of the beam $v_b=0.9c$, where $c$ -- speed of light) with relative density $n_b=0.005n_0$, where $n_0$ -- density of background plasma.  At the current stage of research, it is of interest to assess in principle the influence of large-scale positive and negative density gradients on the beam relaxation process in a model with long term beam injection. The subject of our interest is the change in the beam instability increment, the shift in the localisation region of plasma oscillations, and the dynamics of hydrogen ions with a real mass ratio.

The article is organized as follows. Section \ref{sec:Par} contains description of the particle-in-cell code and computational domain. Parameters of the beam-plasma system are formulated and the increment of the beam instability is analyzed in the linear approximation. In Section \ref{sec:Res}, results of quasi 1D PIC simulations of beam injection into plasmas with homogeneous density profile and descending/rising parabolic gradient are presented. In Section \ref{sec:Conc}, obtained results are discussed and the conclusion is given.

\section{Simulation setup}\label{sec:Par}

For numerical simulations, we use our own 2D3V Cartesian parallel code implemented for Nvidia GPGPU \citep{Lindholm2008}. It is based on standard computational schemes: the  \cite{Yee1966} solver of Maxwell equations for EM fields, the \citet{Boris1970} scheme for solving the equation of motion for collisonless macro-particles with a parabolic form factor, and the charge-conserving \citet{Esirkepov2001} scheme  for calculations of currents.

The computational schemes used allow one to simulate the self-consistent dynamics of hot multicomponent collisionless plasma in electromagnetic fields taking into account relativistic effects. The chosen computational time and space steps allow us to describe both plasma oscillations and rotation of electrons and ions in an external magnetic field. Thus, in such an approach we can consider not only processes of development of kinetic instabilities in plasma and evolution of plasma oscillations, but also to investigate processes of generation of EM radiation of different nature.

In this work we use the model of plasma with open boundary conditions and continuous injection of beam particles. From a technical point of view, various implementations of open boundary conditions are possible, which, as a rule, are present in large computer systems for plasma simulation. As an example, recently the problem of radiation generation by counter propagating electron beams continuously injected at an angle  to each other \citep{Kumar2022} has been studied using the widely used EPOCH code \citep{Arber2015a}. We use self-consistent approach based on the assumption about a slight difference in the plasma parameters in the boundary cells of the computational domain \citep{Annenkov2018}.

Further we will operate with dimensionless values. So particle speeds will be calculated in  units of the speed of light $c$; particle density in units of $n_0$; frequencies in units of the plasma frequency $\omega_{p0}=\sqrt{4\pi e^2 n_0/m_e}$, where $e$ and $m_e$ -- charge and mass of an electron; lengths in $c/\omega_{p0}$; time in the inverse plasma frequency $\omega_{p0}^{-1}$; wavenumbers in $\omega_{p0}/c$; electromagnetic fields in $m_ec\omega_{p0}/c$.   The external magnetic field $B$ will be defined by the dimensionless cyclotron frequency of the electrons $\Omega_e/\omega_{p0}$, where $\Omega_e={eB}/{(m_e c)}$. Thus, by setting $n_0$ we define the main characteristics of the system (frequencies, spatial dimensions, field amplitudes, etc.). At the same time, it is necessary to take into account that it will be correct to use the obtained results only in the range of densities at which (taking into account the temperature of the plasma components) the plasma can be considered as collisionless at the simulation times. This approximation is successfully carried out for solar coronal plasma \citep{Aschwanden2006}.

For information, we give in dimensional form the values at a density $n_0\approx1.24\cdot10^{10}$~cm$^{-3}$, which corresponds to frequency $f={\omega_{p0}}/{2\pi}=1$~GHz. In this case, the unit of distance $c/\omega_{p0}\approx4.8$~cm, the unit of time $\omega_{p0}^{-1}\approx0.16$~ns, the dimensionless electric field amplitude $\widehat{E}=0.2$ corresponds to $E\approx20$~kV/cm, the cyclotron frequency $\Omega_e=0.2\omega_{p0}$ corresponds to magnetic field $B\approx70$~Gs.

\begin{figure}
	
	\includegraphics[width=\linewidth]{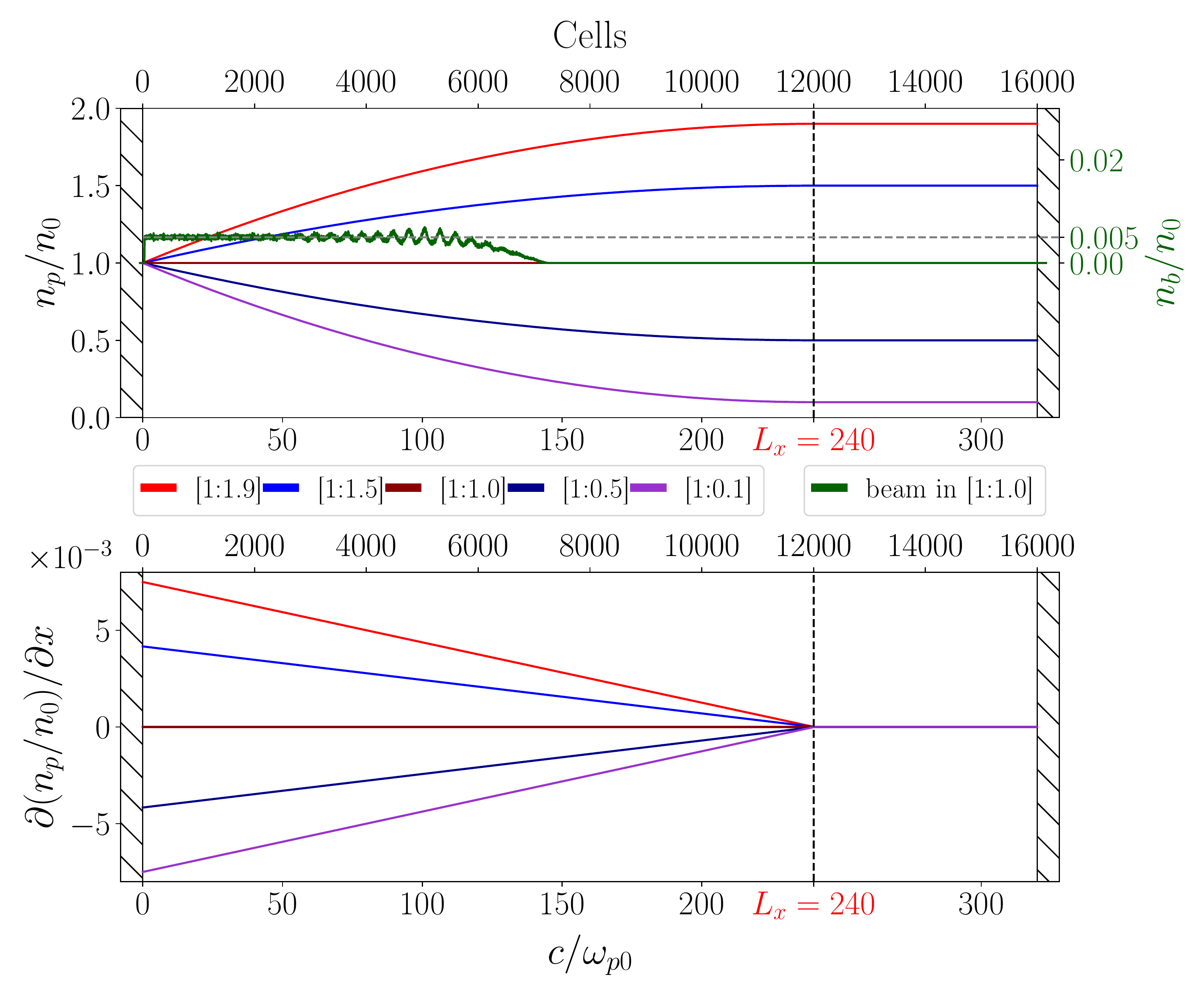}
	\caption{Schematic of the computational region. Top: plasma density profiles and beam densities shortly after the start of injection in uniform plasma. Bottom: longitudinal derivatives of the plasma density profiles.}
	\label{fig:scheme}
\end{figure}

The scheme of the computational domain is shown in figure \ref{fig:scheme}. Plasma buffers implemented open boundaries are located on the ends of the system. The total length of the column of fully ionized hydrogen plasma is $16000$ cells of the size $\Delta x=0.02$ $c/\omega_{p0}$. The time step is $\Delta t=0.01$ $\omega_{p0}^{-1}$, with the total duration  $1000$ $\omega_{p0}^{-1}$.  The length of the area occupied by the gradient is $L_x=240$ $c/\omega_{p0}$ followed by a homogeneous section of plasma. At the left boundary, the initial plasma density is fixed at the value of $n_p=1\cdot n_0$. In the simulations, each particle sort (plasma electrons and ions, beam electrons) is stored as separate arrays, allowing an  independent diagnostic of their parameters at any given time.

In this work, we will consider 5 cases of different densities at the right boundary: $n_1=[1.9;1.5;1;0.5;0.1]$ (Fig. \ref{fig:scheme}). Thus, we consider pairs of increasing and decreasing gradients with the same absolute values of  density derivatives on the coordinate. The change in density in all cases follows the parabolic law. The initial temperature of the plasma electrons is 13 eV. Ions of plasma (protons with a real mass ratio $m_i=1836m_e$) are cold. In the transverse direction, computational domain has the size of 4 cells and periodic boundary conditions. So in fact we are working in one-dimensional mode. This mode is sufficient to investigate the influence of density gradients on the beam relaxation in the case of dominating purely longitudinal instability and allows us to save computational resources and increase the longitudinal length of the region. The latter is important when simulating beams with low relative densities and, accordingly, longer relaxation lengths. The whole system is immersed in an external homogeneous magnetic field $B_x$, such that the electron cyclotron frequency is $\Omega_e=0.2\omega_{p0}$. The presence of a magnetic field contributes to the excitation of purely longitudinal plasma oscillations.  The beam particles are continuously injected through the left boundary. The momentum of each particle is set in accordance with the shifted Maxwell distribution $f_b(\mathbf{p})\propto\exp\left(-(\mathbf{p}-\mathbf{p}_0)^2/\Delta p_b^2\right)$ with the temperature  $T_b=\Delta p_b^2/(2m_e)=13$ eV and directed momentum $\mathbf{p_0}=(p_0,0,0)$ corresponding to the speed $v_b/c=0.9$ (energy $\approx 660$ keV). In addition to the energy dispersion, the beam has a pitch-angle distribution $f_b(\theta)\propto\exp\left(-\theta^2/\Delta \theta^2\right)$ with a spread corresponding to $\Delta \theta=5$ degrees. We consider a basically mono-energetic beam distribution with a small angular spread, since this is sufficient for a first investigation of the principal effect of large-scale gradients. The beam density is  $n_b=0.005n_0$, which is higher than typical beam densities in the region of radio burst generation. The choice of such densities allows one to accelerate the development of beam instability, to reduce the length of beam relaxation and, thus, to save computational resources. A sharp beam profile creates a seed for build-up of longitudinal waves and significantly changes the dispersion of oscillations excited in the plasma. Therefore, in our simulation the beam current grows to a given value in time $50\omega_{p0}^{-1}$. Longer current rise times will lead to an unnecessary increase in the computational complexity of the problem and will also make it impossible to relate the predictions of linear theory for the chosen system parameters to the results of numerical simulations of the beam injection. The main parameters of all calculations are given in the table \ref{tab:params}.

\begin{table}[h]
	\centering
	\caption{Plasma and beam parameters. $n_1$ denotes the density at the right edge of the system.}
	\begin{tabular}{|l|l|l|l|l|l|l|}
		\hline
		run  & $n_1/n_0$ &$v_b/c$   & $n_b/n_0$ & $T_b$, eV & $\Delta\theta$&   $T_e$, eV  \\
		\hline
		run 1 &	1.9 &\multirow{5}{*}{$0.9$}& \multirow{5}{*}{$0.005$}        & \multirow{5}{*}{$13$} & \multirow{5}{*}{$5^{\circ}$ }   &      \multirow{5}{*}{$13$}\\
		run 2 &	1.5& &     &          &    &   \\
		run 3 &	1& &       &          &   &   \\
		run 4 &	0.5& &       &           &    &   \\
		run 5 &	0.1& &         &          &    &   \\
		\hline
	\end{tabular}
	\label{tab:params}
\end{table}

\begin{figure}[h]
	\centering
	\includegraphics[width=\linewidth]{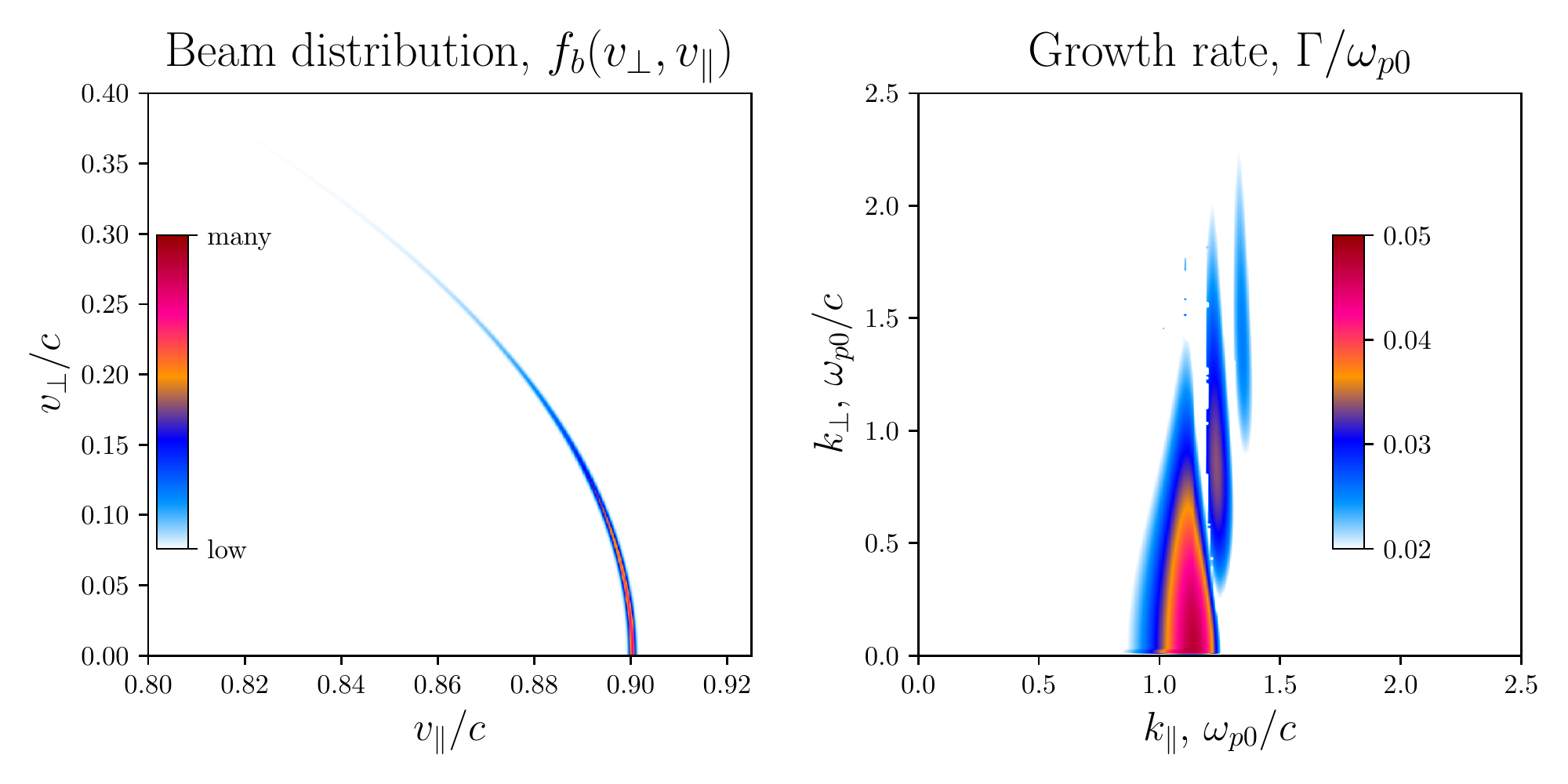}
	\caption{Left: Beam particle distribution function $f_b(v_\perp,v_\parallel)$. Right: the unstable oscillation increment map $\Gamma(k_\perp,k_\parallel)$ of the beam-plasma system in the linear approximation.}
	\label{fig:distInc}
\end{figure}

In Fig. \ref{fig:distInc} left,  the beam particle distribution function is shown. The map on the right shows the increment of a linear instability of oscillations in a homogeneous infinite beam-plasma system, calculated using the numerical library DispLib \citep{Annenkov2021} within the framework of the exact relativistic theory considering a finite magnetic field \citep{Timofeev2013b} in the linear approximation. At the linear stage of instability, the beam excited plasma oscillations can be represented as a plane wave of the form \Mark{$E_x\sim\exp[-\mathrm{i}\omega t]$, where $\omega=\omega_b+\mathrm{i}\Gamma$}, $\omega_b\approx\omega_p$ is the oscillation frequency, and $\Gamma$ is the beam instability increment. For the selected parameters, theory predicts the excitation of purely longitudinal oscillations with near-plasma frequency and longitudinal wave number $k_\parallel\approx 1.13$. The $\Gamma$ value also allows us to estimate the length $L_R=3v_b/\Gamma$ (relaxation length) required for beam trapping. For our beam and plasma parameters $L_R\approx57$ $c/\omega_{p0}$ ($\Gamma\approx0.047$ $\omega_{p0}^{-1}$). For all sorts of particles (plasma electrons, ions and beam electrons) we use $625$ of macro-particles in the cell. The convergence on the number of particles has been confirmed by convergence to a single value of the beam instability increment at the linear stage in calculations with increasing number of particles in the cell in a homogeneous plasma. The achieved value was about 5\% below the predictions of the exact linear theory for an infinite plasma.

Note that in this paper we do not set out to study the parameters of a particular physical system and want to investigate the fundamental influence of plasma density gradients on the relaxation process of continuously injected electron beams. Beam-plasma systems close to the chosen parameters may occur during various high-energy processes on the Sun leading to the generation of fluxes of accelerated electrons propagating along magnetic lines in the solar atmosphere.

\section{Simulation results}\label{sec:Res}

Fig. \ref{fig:Uni} demonstrates the simulation results of beam injection into a homogeneous plasma at different simulation times. Fig. \ref{fig:integ} shows the simulation results for all calculations. On the left sublots the dependence on the coordinate of the longitudinal electric field averaged over the local period of plasma oscillations is shown. The central subplots show the spectra $n_i(q,t)$ of the ion density as a function of time (q is the normalized parallel component of wave number), and the right subplots show the electric field spectra $E_x(\omega,k_\parallel)$. The pink lines indicate the boundaries $v_{ph}=\omega/k=1$. The oscillations inside these boundaries have a superluminal phase velocity $v_{ph}$ and can excite electromagnetic radiation escaping from the plasma. Hereinafter, all spectra and particle distributions are analyzed in the area covered by the gradient ($x<L_x$). Figs. \ref{fig:LMM} and \ref{fig:LMT} show the last moments of each simulation. On the subplots at the top: plasma ion density $n_i(x)$ and their spectrum, on the subplots at the bottom: plasma and beam electron distribution functions $f(v_\parallel,x)$ and $f(v_\perp,v_\parallel)$, where $v_\parallel\equiv v_x$ and $v_\perp=\sqrt{v_y^2+v_z^2}$.  Videos of this data, as well as of the longitudinal electric field, are available in the supplementary materials. Fig. \ref{fig:distr} shows the plasma electron distribution functions $f_e(v)$ and beam $f_b(v)$ with respect to velocity modulus.

\subsection{Uniform density}\label{sec:ResUni}

\begin{figure}
	\includegraphics[width=\linewidth]{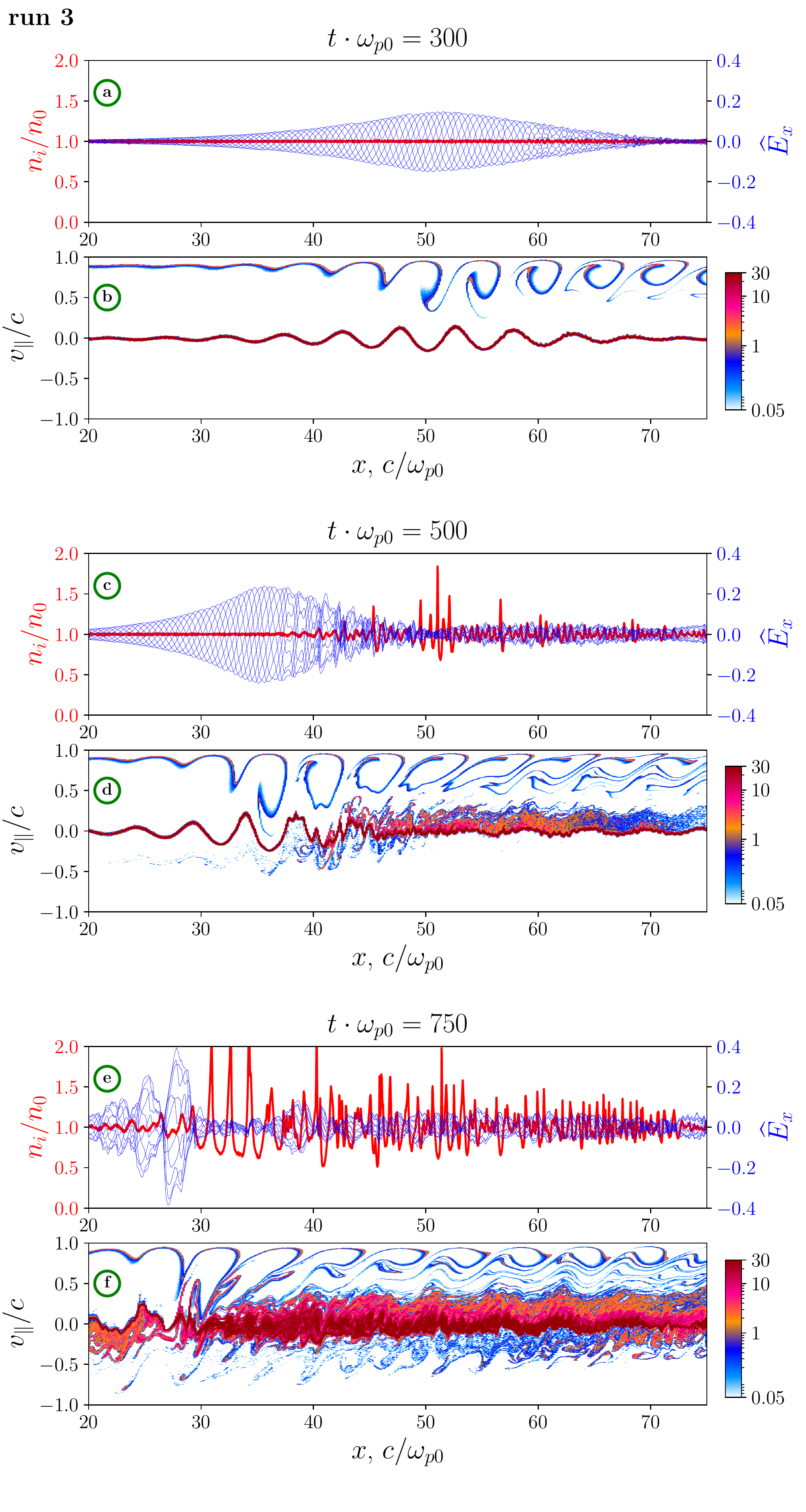}
	\caption{Beam relaxation in a homogeneous plasma at different points in time. (a), (c), (e) Ion density profile (red curves) and longitudinal electric field  $E_x$ (blue curves). Different phases of one oscillation period are shown. (b), (d), (f) Phase portrait $f(v_\parallel,x)$ of beam and plasma electrons.}
	\label{fig:Uni}
\end{figure}

\begin{figure*}
	\includegraphics[width=\linewidth]{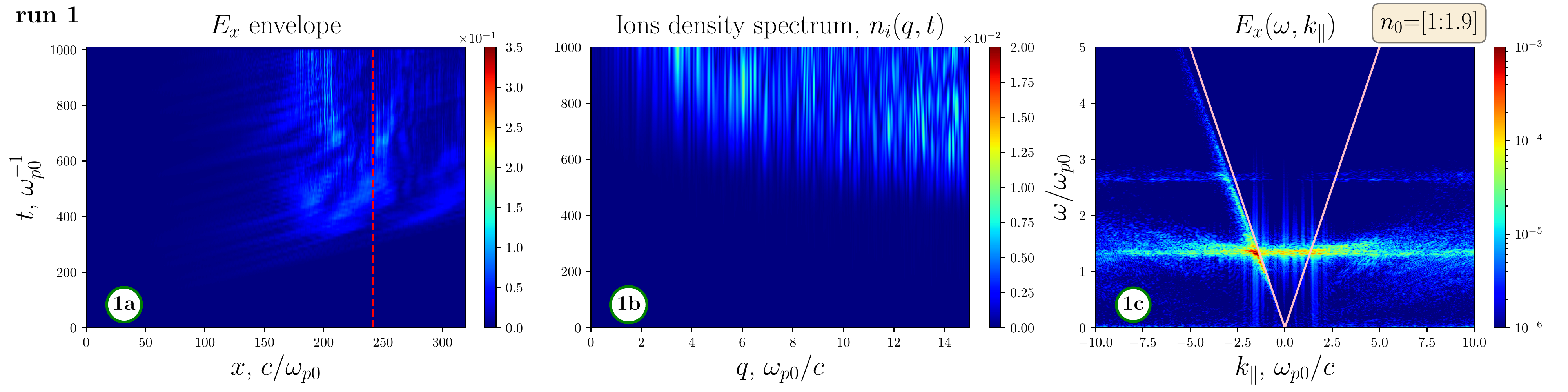}
	\includegraphics[width=\linewidth]{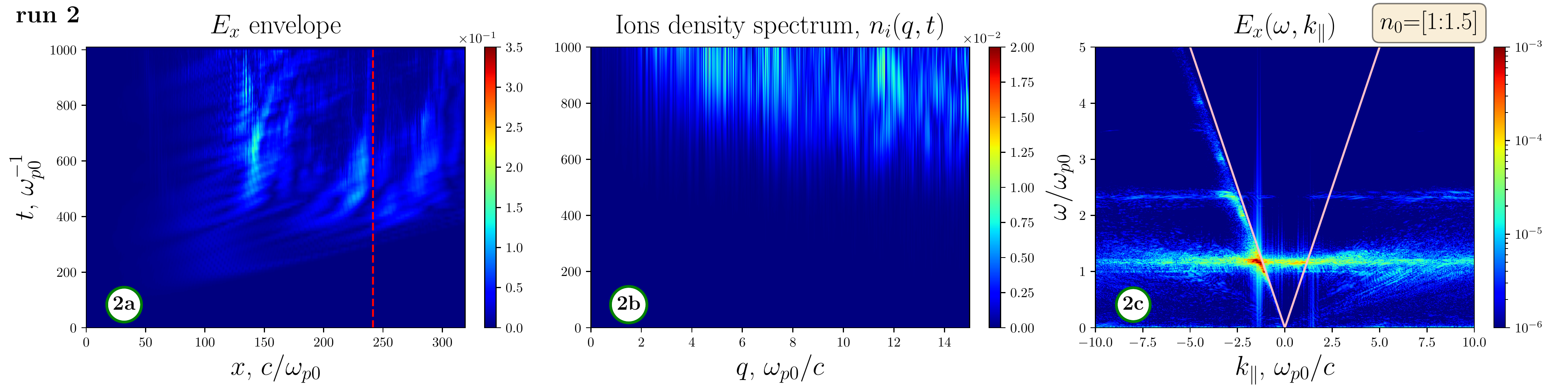}
	\includegraphics[width=\linewidth]{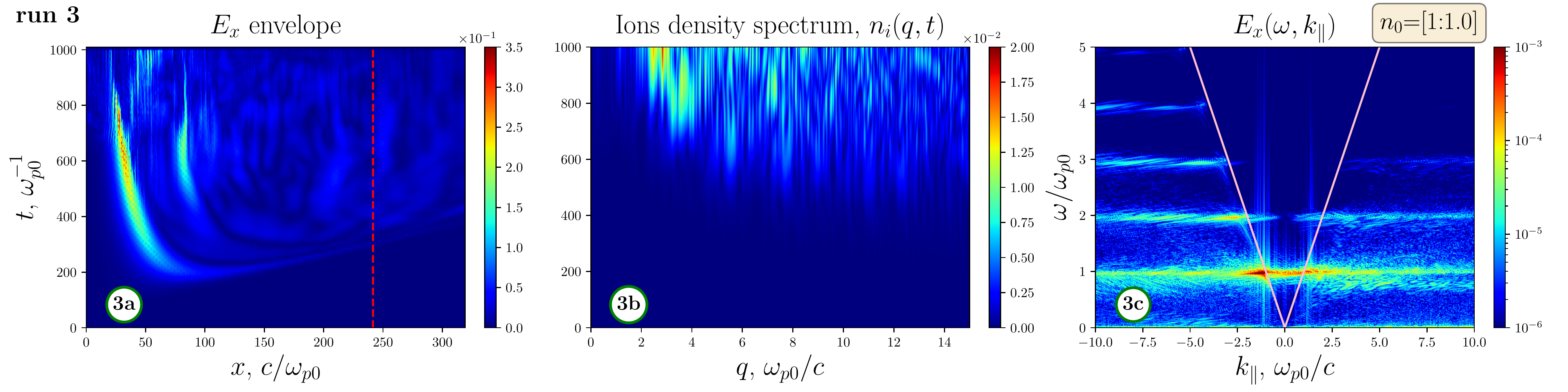}
	\includegraphics[width=\linewidth]{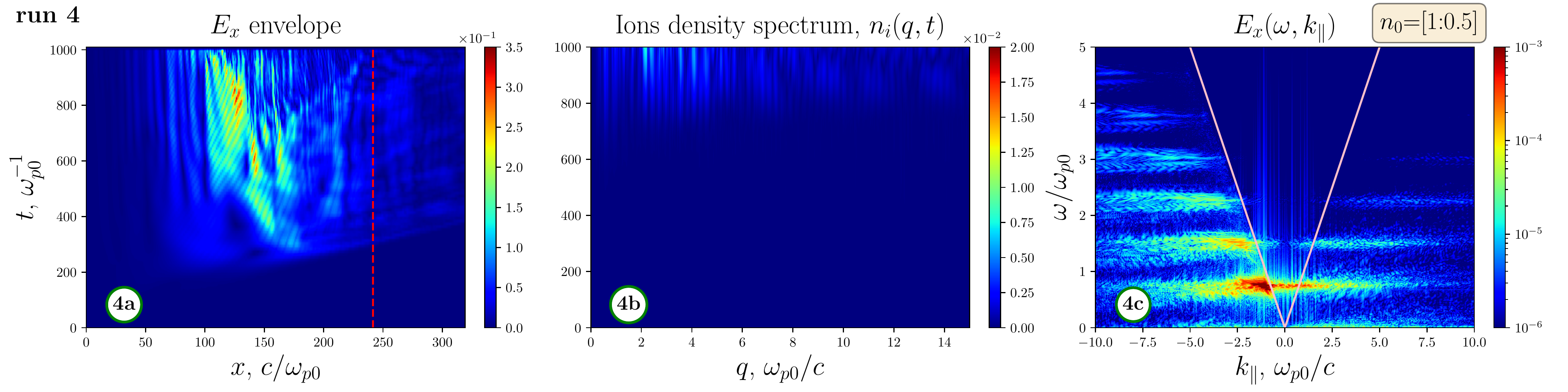}
	\includegraphics[width=\linewidth]{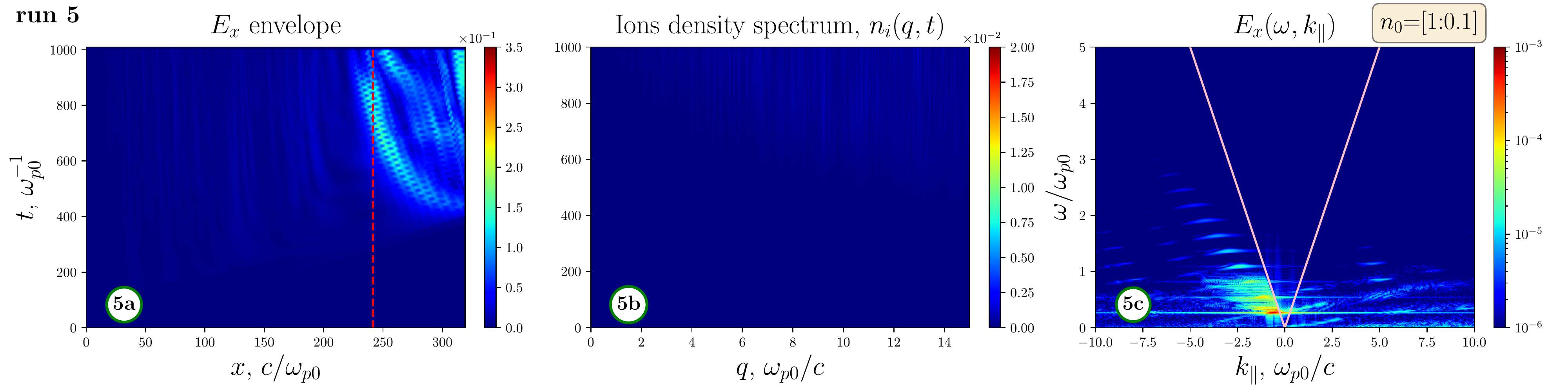}
	\caption{\#a) the longitudinal electric field $E_x$ averaged over the local period of plasma oscillations. Red line indicates gradient edge $L_x=240$ $c/\omega_{p0}$; \#b) ion density spectrum in the gradient area; \#c) the spectrum of the $E_x$ field in the gradient region. The pink lines indicate the boundary $v_{ph}=\omega/|k|=1$.}
	\label{fig:integ}
\end{figure*}

\begin{figure*}
	\includegraphics[width=\linewidth]{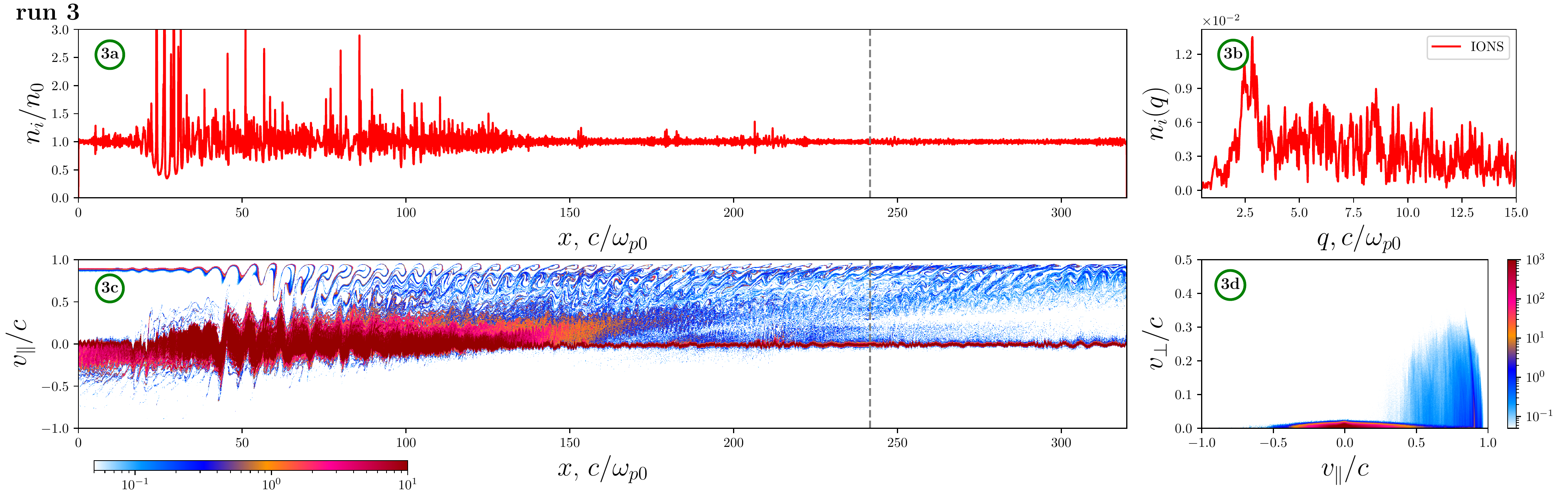}
	\includegraphics[width=\linewidth]{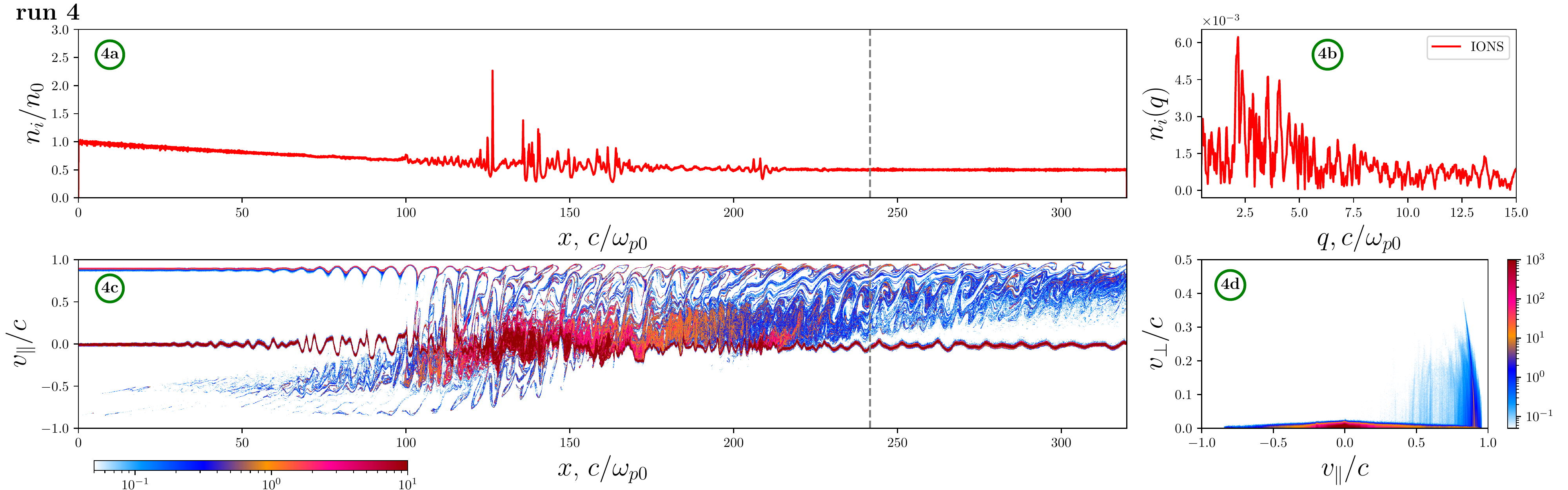}
	
	\caption{The time moment $t\cdot\omega_{p0}=1000$. \textbf{Run 3} and \textbf{run 4}. The regime of modulation instability dominance. \#a) plasma ion density; \#b) ion density spectrum in the gradient region; \#c) phase portrait $f(v_\parallel,x)$ of beam and plasma electrons; \#d) phase portrait $f(v_\perp,v_\parallel)$ of beam and plasma electrons. The gray vertical line marks the boundary of the $L_x=240$ $c/\omega_{p0}$ gradient region.}

	\label{fig:LMM}
\end{figure*}

Let us discuss firstly the relaxation of the beam in a homogeneous plasma (run 3). Due to the small spread in the beam velocity, the development of the beam-plasma instability occurs in a hydrodynamic regime. In fact, the entire beam participates in the excitation of plasma oscillations with a single wave number via the Cherenkov resonance. In the linear phase of the instability (approximately at  $t=200$ $\omega_{p0}^{-1}$), oscillations with a wave number of $k_\parallel\approx1.13$ are excited, which is consistent with the predictions of linear theory for infinite plasma. As can be seen from Fig. \ref{fig:integ} (3a), already after $200$ $\omega_{p0}^{-1}$ two-stream instability develops at a distance of about $50$ $c/\omega_{p0}$ from the injection region, which agrees well with the simple estimate $L_R=3v_b/\Gamma$ made earlier. Further, the beam relaxation region shifts even closer to the injector, forming a main wave packet of large amplitude and a secondary relaxation region at a distance of about $80$ $c/\omega_{p0}$. The linear stage of the two-stream instability is followed by the nonlinear process of beam trapping by the field of the most unstable plasma wave. As a result, spatially localized wave packets with a large amplitude of the longitudinal electric field are formed (Fig. \ref{fig:Uni}a). By this point, the previously excited plasma oscillations are having a significant effect on the beam distribution function. Due to the considerable decrease in the average longitudinal velocity of the trapped beam, the spectrum of pumped beam oscillations shifts to higher wave numbers, so that at $t=300$ $\omega_{p0}^{-1}$ oscillations with $k_\parallel\approx1.2$ dominate in the spectrum. During the beam trapping stage, characteristic circular patterns (Fig. \ref{fig:Uni}b) are formed on the beam phase plane $f(v_\parallel,x)$ whose size directly depends on the amplitude of the plasma wave with which the beam interacts. More details on the theory of electron beam relaxation in the trapping mode can be found in \citet{Timofeev2006} and references therein. When the amplitude of plasma oscillations reaches a sufficiently large value, modulation instability develops \citep{Vedenov1964,Zakharov1972,Thornhill1978}. During its development, the pressure of high-frequency oscillations localized in space leads to the formation of plasma ion density wells (Fig. \ref{fig:Uni}c at $x=50$ $c/\omega_{p0}$). The plasma oscillations are trapped in the forming density wells. These oscillations fall out of the resonance with beam particles, so with time they decay, losing energy to deepening density wells. The oscillations trapped in the density wells can be seen very well in Fig. \ref{fig:Uni}e at $x<30$ $c/\omega_{p0}$.  These density wells have a longitudinal dimension comparable to the plasma wavelength, so growing density gradients as they deepen significantly weakens the beam-plasma instability even to full disruption. Such wells can be seen in Fig. \ref{fig:Uni}e in the region $x>30$ $c/\omega_{p0}$. The formation of such sharp gradients near the injection region causes the beam begins to excite instability at a larger distance from the injection region and the scenario described earlier is repeated. This process can be seen more clearly in the attached video, as well as in Fig. \ref{fig:LMM} for time moment $t=1000$ $\omega_{p0}^{-1}$. The subplot (3a) of Fig. \ref{fig:LMM} shows the density wells formed at less than $40$ $c/\omega_{p0}$ at this time point. At the same time, as can be seen from the beam phase portrait on the subplot (3c), there is no significant beam relaxation in this region, which begins after about $50$~$c/\omega_{p0}$.

This scenario of relaxation of continuously injected electron beams is quite typical. We observed it in the studies of electromagnetic emission processes from the beam-plasma systems for both relativistic \citep{Annenkov2016b} and weakly relativistic beams \citep{Annenkov2019}. The wavelength of the trapped oscillations is automatically adjusted to the modulation length, which allows the beam-plasma antenna \citep{Annenkov2016a} to radiate in the transverse direction at the local plasma frequency. The scattering of the travelling plasma wave on the formed longitudinal density modulation can also lead to the generation of radiation at the second harmonic of the plasma frequency by the beam-plasma antenna mechanism.

\subsection{Density gradients}\label{sec:ResGrad}
\begin{figure}
	
	\includegraphics[width=\linewidth]{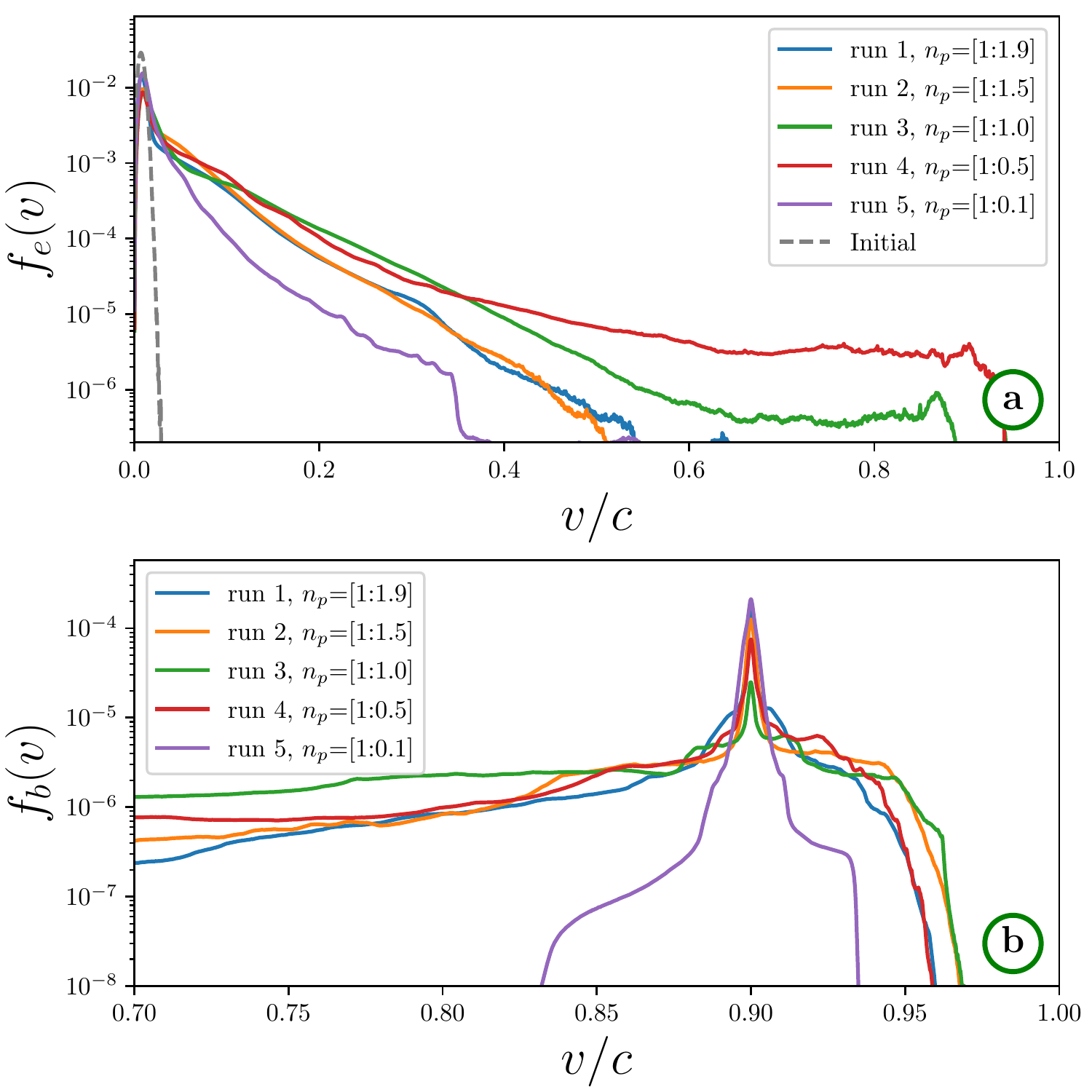}
	\caption{Averaged over the longest period of plasma oscillations, the velocity distribution of (a) plasma electrons (normalized to $1$) and (b) beam electrons at the end of the calculations.}
	\label{fig:distr}
\end{figure}

Consider the effect of density gradients. Let us start with the weakly decreasing gradient mode up to $n_1=0.5n_0$ (run 4). As can be seen from Fig. \ref{fig:integ} (4a), the presence of the gradient leads to some suppression of the instability and no beam relaxation is observed near the injection region. Later, however, the beam with a nearly initial distribution enters a region of lower density and the relative density of the beam $n_b/n_p$ grows, which contributes to the development of instability. In the competition between the gradient suppressing instability and the relative beam density stimulating its growth, the latter wins. After $100$ $c/\omega_{p0}$ a process of intense excitation of plasma oscillations, development of modulation instability and formation of density wells is observed. The depth of density modulation by the end of the simulation turns out to be less than in the case of homogeneous plasma. This is due to the fact that the development of instability in this more distant region from the injector began later. At longer simulations we should expect formation of deeper density wells. The dimensionless amplitude of plasma oscillations in the considered case is approximately the same as in the case of homogeneous plasma ($|E_x|\approx 0.2$). At the same time, the local plasma density is already almost two times less than in the case of homogeneous plasma. For this reason, plasma oscillations have a much greater effect on the plasma electrons, leading to acceleration of the plasma electrons to velocities on the order of the beam speed (Fig. \ref{fig:distr}), but not only in the direction of beam propagation, but also in the opposite direction (Fig. \ref{fig:LMM} 4c). This fact means that the acceleration of the plasma electrons occurs actually during half a period of plasma oscillations, when the electric field can have both positive and negative sign. Thus, the unstable region becomes a source of two flows of electrons with relativistic velocities. The electrons accelerated in the positive direction actually become particles of the initial beam.

The considered regimes with homogeneous plasma and a weak decreasing gradient will be called modulation instability dominance regimes. They are characterized by the excitation of plasma oscillations with an amplitude large enough to develop modulation instability and form long-wave density disturbances (Figs. \ref{fig:LMM} 3b and 4b), in which plasma oscillations can be captured. Such a regime is favorable for activation of the beam-plasma antenna mechanism and generation of radiation at harmonics of the plasma frequency. Large amplitude of plasma waves also leads to nonlinear excitation of many harmonics of plasma frequency (Fig. \ref{fig:integ} 3c and 4c).

\begin{figure}
	\includegraphics[width=\linewidth]{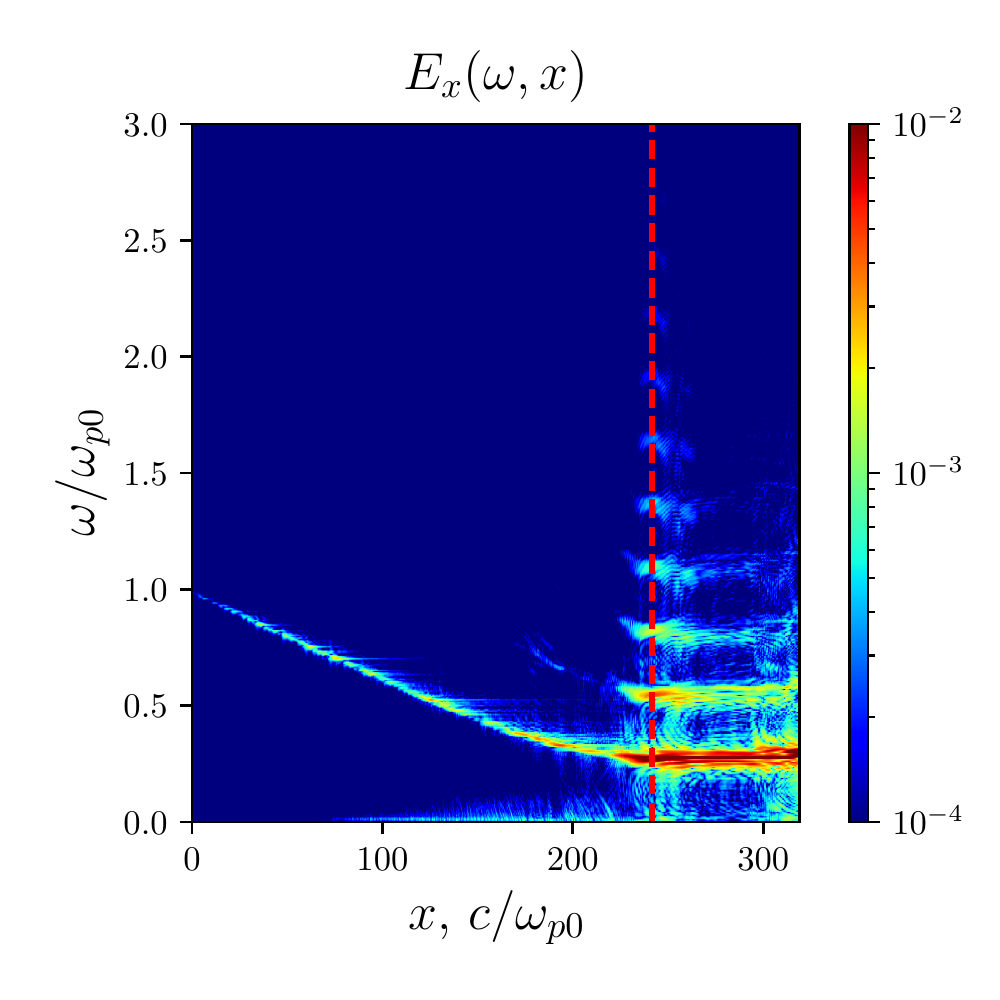}
	\caption{The spectrum of the longitudinal electric field as a function of the coordinate for \textbf{run 5} ($n_0=[1:0.1]$). }
	\label{fig:sp_run5}
\end{figure}

\begin{figure*}
	\includegraphics[width=\linewidth]{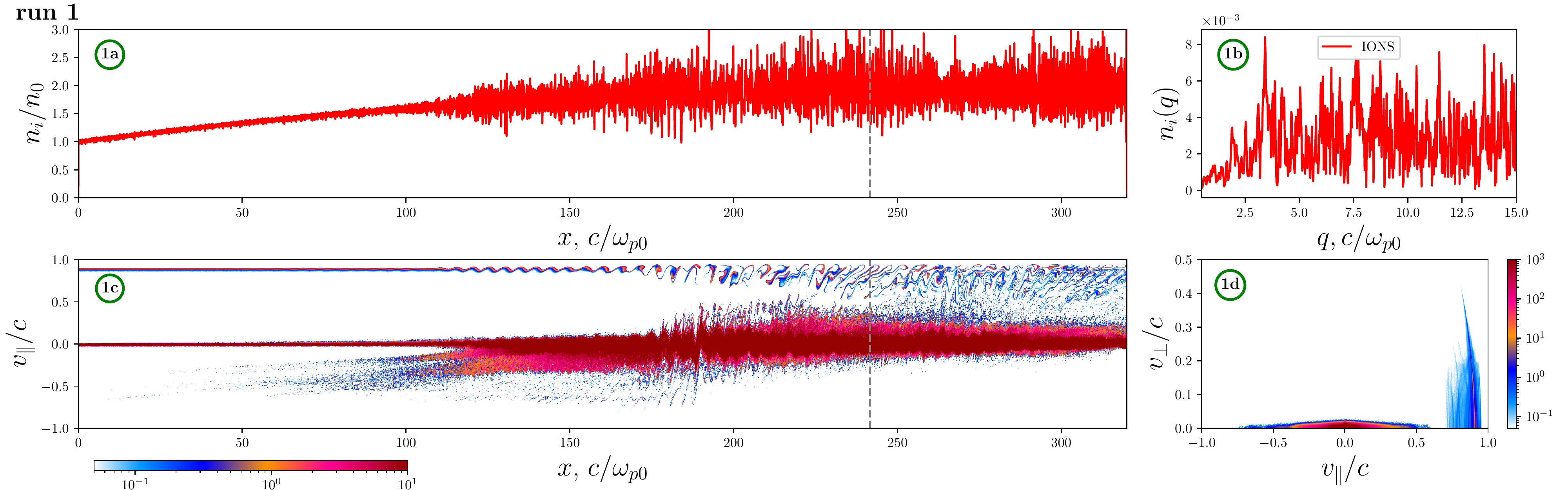}
	\includegraphics[width=\linewidth]{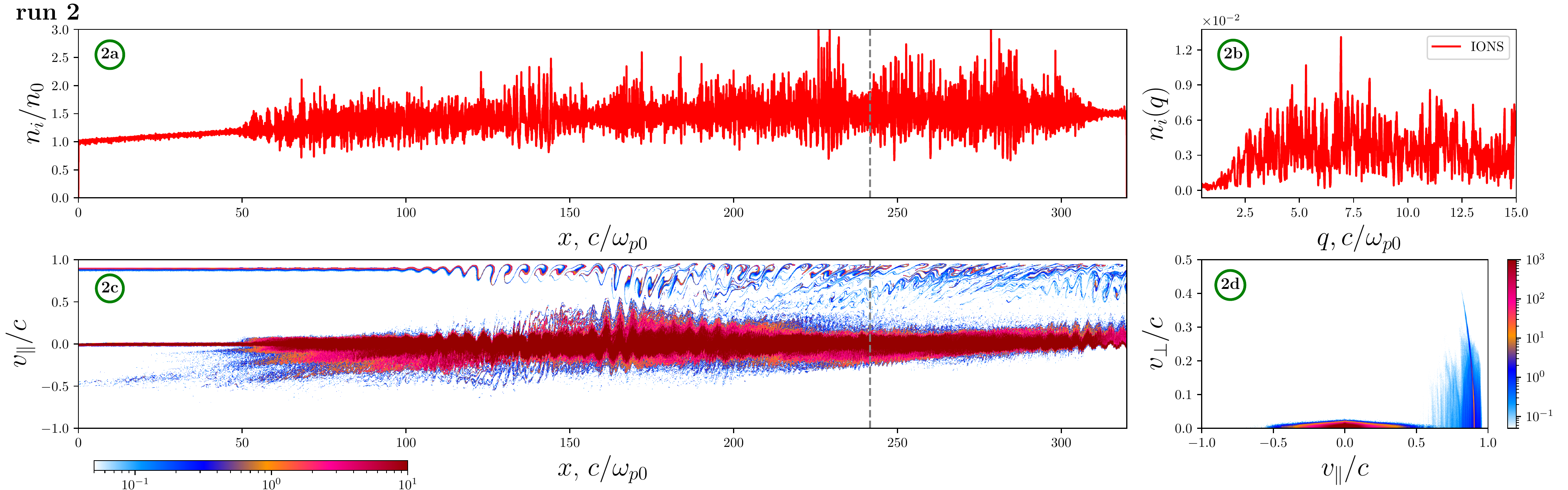}
	\includegraphics[width=\linewidth]{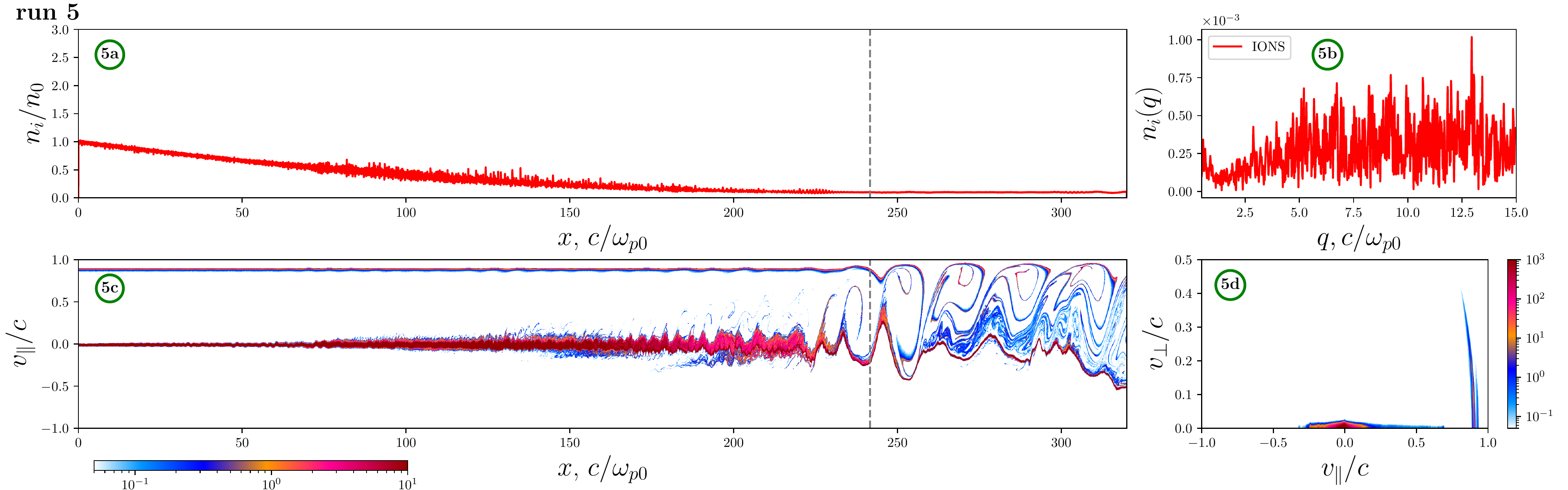}
	\caption{The time moment $t\cdot\omega_{p0}=1000$. \textbf{Run 1}, \textbf{run 2} and \textbf{run 5}. Turbulent mode. \#a) plasma ion density; \#b) ion density spectrum in the gradient region; \#c) phase portrait $f(v_\parallel,x)$ of beam and plasma electrons; \#d) phase portrait $f(v_\perp,v_\parallel)$ of beam and plasma electrons. The gray vertical line marks the boundary of the $L_x=240$ $c/\omega_{p0}$ gradient region.}
	\label{fig:LMT}
\end{figure*}

Let us discuss the other modes, which we will call turbulent ones. As can be seen from Figure \ref{fig:integ} (5a), the greater density decrease in this case dominates in terms of suppression of the beam-plasma instability in comparison with the process of growth of the relative beam density. As the beam propagates along the gradient, it excites plasma oscillations of small amplitude, but as it approaches and passes to a region of homogeneous plasma, a rapid excitation of plasma oscillations of large amplitude occurs (Fig. \ref{fig:LMT} 5c). It can be seen from Fig. \ref{fig:sp_run5} that there is excitation of multiple (up to 10) plasma frequency harmonics in the region of transition to uniform density, which can also be seen in the spectrum of $E_x(\omega,k_\parallel)$ in subplot 5c on Fig. \ref{fig:integ}. In this mode, the least heating of the plasma is observed (Fig. \ref{fig:distr}), and noticeable perturbations of the ion density do not have time to form during the computation time.


At propagation into growing density the effect of stimulation of instability due to growth of relative density of the beam disappears, which on the contrary begins to decrease. Therefore, for run 2, which has the same density change rate as run 4, we do not observe the excitation of oscillations of large amplitude, although the localization of the main wave packet approximately coincides with the case of the decreasing gradient in run 4 (Fig. \ref{fig:integ} 2a and 4a). At more rapid density growth in run 1, the weakening of the instability level continues, which can be seen not only in the reduction of the oscillation amplitude (Fig. \ref{fig:integ} 1a), but also in the reduction of the trapping region (Fig. \ref{fig:LMT} 1c). In these modes, the development of modulation instability and the formation of density wells is not observed due to the small amplitude of the plasma oscillations. 

Let us compare separately the modes with the highest density gradients (increasing run 1 and decreasing run 5). Although, the magnitudes of the density gradients are the same, at the increasing gradient there is more significant excitation of oscillations, as well as greater heating of the plasma (Fig. \ref{fig:distr}a) and diffusion of the beam electron distribution (Fig. \ref{fig:distr}b). The main reason for these differences seems to be the fact that the plasma wave increases its phase velocity as it moves towards the increasing density gradient. Because of this, the wave becomes capable of interacting with faster particles, resulting in greater particle velocity diffusion \citep{Ryutov1969}.

Let us compare the beam and plasma electron distribution functions formed by the end of the simulation. Fig. \ref{fig:distr} b shows the dependences $f_b(v)$. Since all particles in the gradient region are diagnosed, the beam particles just injected into the system with the main velocity $v_b=0.9c$ are clearly visible. It is also seen that in the case of initially homogeneous plasma with large amplitude of excited plasma oscillations (run 3) and in the case of $n_p=[1:1.5]$ gradient (run 2) the particles accelerated to approximately the same energy are observed. Thus, most of the accelerated particles have velocities of the order of $\approx0.95c$ ($\approx 1.1$ MeV), and there is also a population with velocities as high as $\approx0.97c$ ($\approx 1.6$ MeV). In calculations with gradients $n_p=[1:1.9]$ (run 1) and $n_p=[1:0.5]$ (run 4), the bulk of the accelerated particles have speeds $\approx 0.94c$ ($\approx 0.99$ MeV) and there is a population of electrons with speeds up to $\approx0.96c$ ($\approx 1.3$ MeV). The formation of regular superthermal tails and the appearance of a significant number of particles accelerated to beam velocities are observed on the plasma electrons. Thus, in the case of a homogeneous plasma (run 3) the characteristic maximum velocity of such particles is $\approx 0.87c$ ($\approx 0.5$ Mev), while for a decreasing gradient $n_p=[1:0.5]$ (run 4) a noticeable fraction of plasma particles are accelerated to velocities $\approx 0.94$ ($\approx 0.99$ MeV). It can be seen that such a scenario of acceleration of plasma particles is specific only for regimes with large amplitude of excited oscillations, when the beam trapping region extends up to the plasma particles. In this case, the wave can, on the one hand, accelerate the plasma particles to the beam energy and, on the other hand, slow down the beam particles and actually turn them into plasma particles.

\begin{figure}
	\centering
	\includegraphics[width=0.8\linewidth]{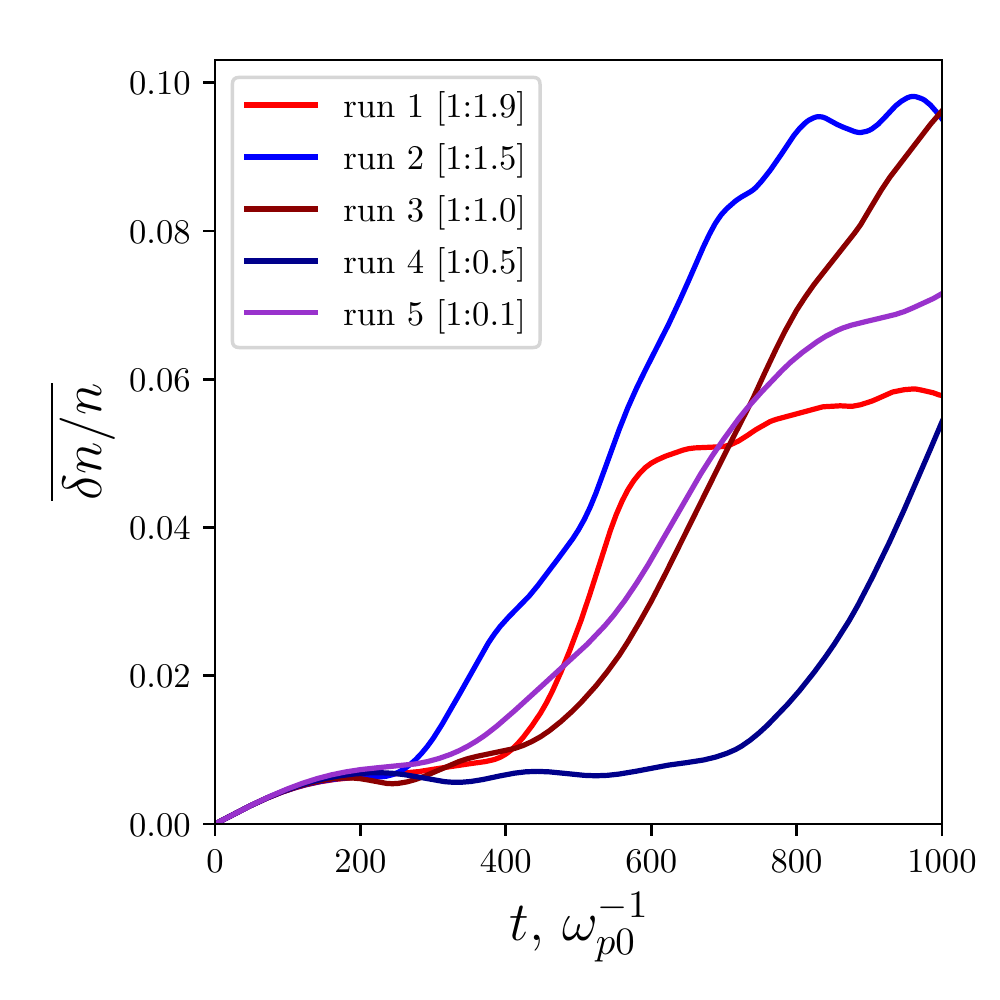}
	\caption{Average ion density fluctuations as a function of time.} 
	\label{fig:fluct}
\end{figure}

Let us try to estimate quantify the level of fluctuations that arise from the instability in the simulations. For this purpose let us plot the time dependence of the average relative value of the ion density perturbations: \begin{equation}
		\overline{\delta n/n}^{(\sigma)}(t)=\left(\sum_{x=0}^{N_x}\frac{|n^{(\sigma)}(x,t)-n_p^{(\sigma)}(x)|}{n_p^{(\sigma)}(x)}\right)/ N_x,
	\end{equation} where $N_x=12000$ is the number of cells occupied by the gradient, $(\sigma)$ marks the different runs, $n^{(\sigma)}(x,t)$ is the plasma ion density in cell $x$ at time $t$, $n_p^{(\sigma)}(x)=n^{(\sigma)}(x,t=0)$ is the plasma density at the initial time. Fig. \ref{fig:fluct} shows the resulting dependencies. Up to about $t=200$ $\omega_{p0}^{-1}$, the dependence is similar for all runs. At this time, density fluctuations related to thermal motion of plasma particles grow. Further growth of density fluctuations is determined by the characteristics of interaction with the electron beam. From the obtained results one can see a different time behavior of this quantity for the modes in which the beam-plasma instability was significantly attenuated by the density gradient (run 1, run 2 and run 5) and the modes with intense beam relaxation and development of modulation instability (run 3 and run 4). In the first case, the value of relative density fluctuations reaches saturation even during the simulation time. At the same time, the final values turn out to be similar in order of magnitude. In the second case, we see approximately the same growth rate of fluctuations and no saturation during the simulation time. We can suppose that further in this mode the average over the system length relative density fluctuations should reach some saturation, since there is a cyclic process of growth of deep density holes due to modulation instability, local failure of the beam-plasma instability in this region and gradual relaxation of density fluctuations to more uniform state, allowing a new development of the beam-plasma instability in the system.

\section{Conclusion}\label{sec:Conc}
In this paper, we have investigated the effect of increasing and decreasing plasma gradients on the relaxation process of a relativistic electron beam with a pitch-angle distribution in an external magnetic field. For the selected parameters of the beam-plasma system, 2 regimes were found.

The first regime is realized for homogeneous plasma and plasma with relatively smoothly decreasing density. In this mode, a well-localized packet of plasma oscillations with a sufficiently large amplitude for the development of modulation instability arises. The pressure of high-frequency oscillations forms density wells and produces longitudinal density modulation. Captured in such wells oscillations can generate radiation at plasma frequency, while travelling oscillations interacting with modulation can excite superluminal perturbations at plasma and doubled plasma frequencies, generating radiation by the mechanism of the beam-plasma antenna.

The second regime is realized in the case of a sharply decreasing gradient and in both simulations with increasing density. In this mode, small-amplitude plasma oscillations are more uniformly distributed along the plasma length and the development of modulation instability is not observed. Instead of long-wave modulation of ion density, a short-wave turbulent spectrum is observed. The small amplitude of oscillations also limits the nonlinear excitation of high harmonics of plasma oscillations. 

In all cases, a population of superthermal electrons is formed on the plasma electron distribution function. However, in modes with large amplitude, acceleration of plasma electrons to energies of beam particles was observed. Moreover, in the case of a smoothly decreasing gradient, a noticeable fraction of the particles were accelerated in the opposite direction from the beam propagation direction. The formation of high-energy populations is also observed on the beam electron distribution function. The most energetic particles were formed in two cases. In the first, there was a relatively smooth growth of the plasma density as the beam moved through the system. The acceleration of particles in this case, apparently, is associated with the classical mechanism of growth of the phase velocity of plasma oscillations during propagation toward the growing gradient, which allows them to interact with faster and faster particles. In the second case, the acceleration of particles took place in an initially homogeneous plasma in which a local wave packet of large amplitude was formed. Thus, the presence of a growing density gradient is not a necessary condition for the appearance of high-energy particles in the case of development of sufficiently strong beam-plasma instability. For calculations with a smooth decreasing gradient and a sharp increasing one, the acceleration of the beam particles is also observed, but to lower values. Apparently, in the case of a smooth gradient we can expect greater acceleration of particles during further development of the instability.

In this work, we have considered a relativistic beam with small energy dispersion, due to which the development of the beam instability occurred in the hydrodynamic regime. It is expected that a small change of basic energy of the beam will not lead to essential difference of the evolution scenario of the system, since the value of the instability increment and the wave number of the excited plasma wave $k_\parallel\approx\omega_{p0}/v_b\approx\omega_{p0}/c$ will not change dramatically. During the transition to ultrarelativistic beams the answer is not so clear, because due to increasing anisotropy of relativistic electron mass of the beam, oblique plasma oscillations (despite the presence of a small magnetic field) may become most unstable. On decreasing the beam energy to sub-relativistic values both the increment of instability and the spectrum of excited oscillations will change essentially. Therefore, in this case, one can expect significant differences from the obtained results. Increasing of energy spreads up to keV scales in the relativistic case does not change essentially the value of the instability increment and should not change essentially the result. In the case of nonrelativistic beams larger energy spreads will lead faster to the transition to the kinetic two-stream instability mode. Also if one sets the beam with large velocity spreads the injection into the plasma at considered times will proceed unevenly: fast particles will rapidly move away from the injector, disturbing the plasma before the arrival of the distribution core, while slow particles will noticeably delay. The whole relaxation process will not go in a very predictable way. Such a problem statement is in principle impossible to consider in a simplified model of an infinite beam-plasma system (because a given beam distribution is set at once over the whole space), but its investigation in a model with an injected beam looks possible. The study of the influence of these and other factors on the electron beam relaxation process in inhomogeneous plasma is of interest for further research.

\section{Acknowledgments}

This work was supported by the grant MK-2676.2021.1.2 by Ministry of Science and Higher Education of the Russian Federation. Simulations were performed using the computing resources of the Center for Scientific IT-services ICT SB RAS.




%

\end{document}